\documentclass[12pt, a4paper]{article}
\pdfoutput=1
\usepackage{graphicx}
\usepackage{amssymb}
\usepackage{amsmath}
\usepackage{bm}
\usepackage{color}
\usepackage{theorem}
\usepackage{subfigure}

\usepackage[sort&compress,numbers, merge]{natbib}

\definecolor{Orange}{cmyk}{0,0.61,0.87,0}
\definecolor{JungleGreen}{cmyk}{0.99,0,0.52,0}
\definecolor{OliveGreen}{cmyk}{0.64,0,0.95,0.40}
\definecolor{Brown}{cmyk}{0,0.81,1,0.60}
\definecolor{RoyalBlue}{cmyk}{0.71,0.53,0,0.12}

\usepackage[colorlinks=true, linkcolor=OliveGreen, citecolor=RoyalBlue,
urlcolor=Brown]{hyperref}

\newcommand{\beq}{\begin{equation}}   
\newcommand{\eeq}{\end{equation}}
\newcommand{\beqn}{\begin{eqnarray}}   
\newcommand{\eeqn}{\end{eqnarray}}

\newcommand*\xbar[1]{%
 \kern0.5ex%
  \hbox{%
   \kern0.2ex%
      \vbox{%
      \hrule height 0.5pt 
      \kern0.5ex
      \hbox{%
        \kern-0.1em
        \ensuremath{#1}%
        \kern-0.1em
      }%
    }%
  }%
}

\newcommand{\gsim}{\lower.7ex\hbox{$
\;\stackrel{\textstyle>}{\sim}\;$}}
\newcommand{\lsim}{\lower.7ex\hbox{$
\;\stackrel{\textstyle<}{\sim}\;$}}
\begin{document}
\begin{titlepage}

\begin{flushright}
UMN--TH--3522/16\\
FTPI--MINN--16/11
\end{flushright}

\vskip 1.35cm
\begin{center}

{\Large
{\bf A Visible QCD Axion from an Enlarged Color Group}}

\vskip 1.5cm
{\bf Tony Gherghetta}$^a$,
{\bf Natsumi Nagata}$^b$,
{\bf Mikhail Shifman}$^{a,b}$

\vskip 0.5cm

{\small
{\it $^a$School of Physics \& Astronomy, University of Minnesota,
 Minneapolis, MN 55455, USA}\\[3pt]
{\it $^b$William I. Fine Theoretical Physics Institute, School of
 Physics \& Astronomy, \\ University of Minnesota, Minneapolis, MN 55455,
 USA}}

\date{\today}

\vskip 1.5cm
\begin{abstract}

We consider the possibility of an enlarged QCD color group, ${\rm
SU}(3+N')$ spontaneously broken to 
${\rm SU}(3)_c\times {\rm SU}(N')$ with extra vector-like quarks transforming
in the fundamental representation. When the heavy quarks are integrated 
out below the PQ-breaking scale, they generate an axion coupling which 
simultaneously solves the strong CP problem for both gauge groups. 
However, the axion mass now receives a  new nonperturbative contribution 
from the ${\rm SU}(N')$ confinement scale, which can be substantially
larger than the QCD scale. This can increase the axion mass to be at or
above the electroweak scale. This visible axion can then decay into
gluons and photons giving rise to observable signals at Run-II of the
LHC. In particular, if the mass is identified with the 750~GeV diphoton
resonance, then the new confinement scale is $\sim \text{TeV}$ and the
PQ-breaking scale is $\sim 10~\text{TeV}$. This predicts vector-like quarks 
and a PQ scalar resonance in the multi-TeV range, with the possibility that 
dark matter is an ${\rm SU}(N')$ baryon.

\end{abstract}

\end{center}
\end{titlepage}

\section{Introduction}
 
It has long been known that a nonzero $\theta$-angle in QCD leads to
large CP-violating effects which are not observed, such as a neutron
electric dipole moment \cite{Crewther:1979pi, Baluni:1978rf}. A simple
way to address this strong CP problem is to introduce a global
Peccei--Quinn (PQ) symmetry \cite{Peccei:1977hh, Peccei:1977ur} which is
spontaneously broken at a scale $f_a$ and gives rise to a
Nambu--Goldstone (NG) boson, the axion \cite{Weinberg:1977ma,
Wilczek:1977pj}. Nonperturbative effects then generate an axion
potential with a minimum that occurs at an axion vacuum expectation
value (VEV) that cancels a nonzero $\theta$-angle, thereby dynamically
solving the strong CP problem. The axion can be considered to be part of
a complex scalar field $\Phi$, which couples to vector-like quarks in
the fundamental representation of the QCD color group ${\rm SU}(3)_c$, and 
is charged under the PQ symmetry~\cite{Kim:1979if, Shifman:1979if}. When $\Phi$ 
obtains a VEV, $\langle \Phi\rangle = f_a/\sqrt{2}$, the PQ symmetry is
spontaneously broken and the vector-like quarks obtain a mass. After
these quarks are integrated out, they generate an axion coupling to the
gluon field strength, giving a simple realization of the PQ mechanism. 

The QCD axion solution relates the axion mass $m_a$ to the PQ-breaking scale $f_a$. 
For example, in the KSVZ model~\cite{Kim:1979if, Shifman:1979if}, the relation, 
assuming two light quarks, is given 
by
\beq
m_a^2 f_a^2  = \frac 1 8\,  f_\pi^2m_\pi^2 \,  \frac{4m_u\,m_d}{(m_u+m_d)^2}\,,
\label{2}
\eeq
implying that 
\beq
 m_a^2 f_a^2 \sim \frac 18\Lambda_c^4.
 \label{2p}
 \eeq
Here $\Lambda_c$ is the QCD confinement scale, and we have used the
experimental values of the quark masses, the pion decay constant $f_\pi
\simeq 130$~MeV and the pion mass $m_\pi \simeq 135$~MeV.  Note that the right-hand side 
of Eq.~\eqref{2}, which is given by the topological susceptibility~\cite{Shifman:1979if} 
\begin{equation}
 {\cal T} \equiv -i \int d^4x~ \langle 0| {\rm T}
\left[
\frac{1}{32\pi^2}G^a_{\mu\nu} \widetilde{G}^{a\mu\nu}(x),~
\frac{1}{32\pi^2}G^b_{\rho\sigma} \widetilde{G}^{b\rho\sigma}(0)
\right]|0\rangle ~,
\end{equation}
tends to zero in the chiral limit, where $G^a_{\mu\nu}$ is the gluon
field-strength tensor and $\widetilde{G}^a_{\mu\nu} \equiv
\frac{1}{2}\epsilon_{\mu\nu\rho\sigma} G^{a\rho\sigma}$ with
$\epsilon^{\mu\nu\rho\sigma}$ the totally antisymmetric tensor
($\epsilon^{0123} =+1$).  The fact that $ f_\pi m_\pi  \sim \Lambda_c^2$
is a numerical coincidence. In the absence of light quarks, the
topological susceptibility is of order $\Lambda_c^4$
\cite{Shifman:1979if} and, therefore, $ m_a^2 f_a^2\sim \Lambda_c^4$.

The electroweak scale would be a natural choice for the value of $f_a$,
as was first considered in Refs.~\cite{Weinberg:1977ma,
Wilczek:1977pj}. However, the Weinberg--Wilczek axion was ruled out almost
immediately, while the current astrophysical and cosmological
constraints on invisible axions~\cite{Kim:1979if, Shifman:1979if} 
restrict $f_a$ to lie in the narrow range $10^9~\text{GeV}
\lesssim f_a\lesssim 10^{12}~\text{GeV}$ (although the upper bound, 
due to dark matter over-closure, can be relaxed if the initial
misalignment angle is tuned). These bounds result from the fact that
using Eq.~\eqref{2} with $\Lambda_c\sim 250$ MeV makes the axion
sufficiently light ($10^{-5}~\text{eV} \lesssim m_a \lesssim
10^{-2}~\text{eV}$) that it can be produced in stars. For instance,
a stringent constraint comes from the observation of supernova
(SN)1987A. The axion emission must not shorten the burst duration
implying $f_a \gtrsim 4\times 10^8$~GeV (see, {\it e.g.},
Refs.~\cite{Kawasaki:2013ae, Kim:2008hd,Peccei:2006as}). Moreover, in
the center of the Sun, keV axions (which were originally predicted with
$f_a\simeq$~electroweak scale) can be produced through the
axion-photon conversion in the presence of the solar magnetic
field. Negative results from searches for such axions lead to a similar
albeit less stringent bound. Clearly to invalidate current astrophysical
and cosmological constraints and allow heavier axion masses with 
electroweak values of $f_a$, the relation (\ref{2}) must therefore be modified. 

To untie the relation (\ref{2p}) between $m_a$ and $\Lambda_c$, we will
entertain the possibility that above some ultraviolet (UV) unification scale, $M_U$, there
is an enlarged QCD gauge group ${\rm SU}(3+N')$, which is then
spontaneously broken as
\beq
{\rm SU}(3+N') \rightarrow {\rm SU}(3)_c\times {\rm SU}(N')\,.
\label{brea}
\eeq
The $\theta$ angle  from the ${\rm SU}(3+N')$ group
descends down to the ${\rm SU}(3)_c$ and $ {\rm SU}(N')$
subgroups intact. The quark fields  at short distances belong
to the fundamental representation of ${\rm SU}(3+N')$, and can be
decomposed with regards to ${\rm SU}(3)_c$ and $ {\rm SU}(N')$,
according to Eq.~(\ref{brea}). 

As in the KSVZ model,  extra vector-like quarks are
charged under a PQ symmetry but now they transform in the fundamental
representation of both ${\rm SU}(3)_c$ and ${\rm SU}(N')$. The PQ
symmetry is spontaneously broken by a complex scalar field $\Phi$ with
the axion identified as the NG boson. The extra vector-like quarks $\Psi$ obtain a
mass, $hf_a$ where $h$ is a Yukawa coupling.
When they are integrated out, they generate an axion
coupling to both gauge field strengths. Since both ${\rm SU}(3)_c$ and
${\rm SU}(N')$ originate from a unified color group ${\rm SU}(3+N')$,
they have the same $\theta$ angle, which is not renormalized at low
energies. The axion coupling to the topological charge in these subgroups 
will be the same too. In addition, since the physical theta parameter is 
$\bar\theta = \theta + \text{arg}(\text{det}{\cal M})$, where ${\cal M}$ is a 
complex mass matrix, unification guarantees the same Yukawa terms and, 
therefore, the same phase $\text{arg}(\text{det}{\cal M})$ in the two sectors.
This assumes that no new phases are introduced when the unified partners 
of the Standard Model quarks are decoupled, and a possible UV framework which 
sequesters the ${\rm SU}(3+N')$-preserving CP violation from the symmetry 
breaking is given in Appendix~\ref{sec:decpart}. Thus, when nonperturbative 
effects generate an axion potential, the axion VEV will simultaneously solve 
both strong CP problems.  

Since the colored matter content of the two groups is not necessarily the same 
(and $N'$ is not necessarily equal to 3), the ${\rm SU}(N')$ group can confine at a
scale $\Lambda' \gtrsim \Lambda_c$. This gives a new contribution to the
axion mass relation which now becomes  
\beq
m_a^2\,  f_a^2 \sim \frac{1}{8} \Lambda_c ^4 +  \Lambda'^4\,,
\label{newmass}
\eeq
where we have assumed that there are no light quarks in the SU($N^\prime$)
sector. A dramatic consequence of the modification (\ref{newmass}) is
that the axion can now have an electroweak scale mass!  

An electroweak scale axion can be searched for at the LHC and future
colliders since the generic signal is decays to photons, gluons, and
possibly $W$ and $Z$ bosons and Standard Model quarks and leptons. 
Not only is this experimentally accessible but it is also theoretically appealing 
because the global PQ symmetry is known to be explicitly violated by 
gravitational effects. In
order not to spoil the PQ mechanism, these gravitational effects must
also be suppressed to a very high order in the case of invisible axion
models \cite{Kamionkowski:1992mf, Holman:1992us, Barr:1992qq}; this
difficulty results from the fact that the PQ-symmetry breaking scale is
very high in these models, and thus the effects of Planck-suppressed
operators are sizable compared to the QCD effects on the generation
of the axion potential. An axion at the electroweak scale
helps to suppress the gravitational violations, without any need for further 
sequestering mechanisms.

In particular, the electroweak axion can be identified with the recent 750~GeV
diphoton resonance \cite{ATLASdiphoton, CMS:2015dxe, ATLASdiphoton2,
CMS:2016owr}. This requires a confinement scale $\Lambda'\sim 1$~TeV and 
a PQ-breaking scale $f_a \sim 10$ TeV. With these values, the PQ scalar radial 
mode and vector-like quarks have masses in the multi-TeV range. Furthermore, 
the required cross section for the diphoton excess can be fit if the vector-like 
quarks have ${\cal O}(1)$ hypercharges.  Thus, an electroweak axion gives 
a simple picture of the putative signal.

The idea of extending the color group to raise the axion mass was first considered 
in Refs.~\cite{Dimopoulos:1979qi,Dimopoulos:1979pp}, where unlike in our case, 
the unified quark partners remain below the symmetry-breaking scale.
A modified axion mass relation (\ref{newmass}) was also proposed by
Rubakov~\cite{Rubakov:1997vp}, who considered a mirror copy of the
Standard Model with gauge group ${\rm SU}(5)\times {\rm SU}(5)$. For
subsequent work, see Refs.~\cite{Berezhiani:2000gh, Gianfagna:2004je,
Hook:2014cda, Fukuda:2015ana, Albaid:2015axa}. More recently this mirror
version was studied in Ref.~\cite{Chiang:2016eav} in order to obtain a
visible QCD axion, which was then used to explain the recent diphoton
excess where the PQ scalar radial mode was identified with the 750 GeV
resonance. The difference with our approach is that we do not require a
mirror copy of the Standard Model. Instead, in our model, the two colored
sectors are related by a unified gauge group with a minimal particle
content. This means that we do not have mirror copies of Standard Model
quarks and leptons which leads to extra collider and cosmological
constraints on the axion sector that results from the more complicated
phenomenology. Furthermore, we identify the 750 GeV resonance with an
axion which directly decays to two photons, as opposed to the PQ scalar
radial mode whose decay via a pair of axions produces a four-photon
signal \cite{Chiang:2016eav}.

\section{Enlarging QCD color}
\label{sec:enqcd}

\subsection{Gauge couplings and vacuum angles}
\label{sec:couptheta}

We will assume that the QCD color group ${\rm SU}(3)_c$ is a subgroup of
${\rm SU}(3+N')$. In the UV, the Lagrangian is given by
\beq
{\cal L}= -\frac{1}{4g^2} F_{\mu\nu}^A\, F^{A\mu\nu}  + \frac{\theta
}{32\pi^2}\,F_{\mu\nu}^A\, \widetilde{F}^{A\mu\nu}\, ,
\label{UVLag}
\eeq
where $F^A_{\mu\nu}$, $g$, $\theta$ are the field strength tensor,
gauge coupling, and $\theta$ parameter of the ${\rm SU}(3+N')$
gauge theory, respectively, and $A$ is the adjoint color index of ${\rm
SU}(3+N')$, $A=1,2, ..., {(3+N')}^2-1$. At a scale $M_U$, this
group is spontaneously broken down to ${\rm SU}(3)_c\times {\rm SU}(N')$
where ${\rm SU}(N')$ is the hidden color gauge group. This occurs via
the VEV of an adjoint Higgs field $\Sigma$:
\beq
\langle \Sigma \rangle = V \,\,\mbox{diag} \bigl\{ N^\prime,N^\prime,N^\prime, \,\,
\underbrace{-{3}, -{3}, ...,-{3}}_{N^\prime}\bigr\} ~.
\label{sigvev}
\eeq
In addition, we require that the U(1) subgroup of ${\rm SU}(3+N')$ is broken
at approximately the same scale $V$, by the ${\rm SU}(3)_c\times {\rm SU}(N')$ singlet
component VEV of a scalar field transforming as a three-index antisymmetric tensor of 
${\rm SU}(3+N')$ with zero hypercharge. After this combined symmetry breaking, the gauge bosons 
not belonging to ${\rm SU}(3)_c\times  {\rm SU}(N')$ acquire masses proportional to $g V$ and 
can be dropped from the sum in Eq.~\eqref{UVLag}. Thus, below the scale $gV$,
the Lagrangian becomes 
\begin{equation}
 {\cal L} = -\frac{1}{4g^2}\left[\sum_{a=1}^{8} G^a_{\mu\nu}G^{a\mu\nu}
+ \sum_{\alpha =1}^{N^{\prime 2}-1} G^{\prime \alpha}_{\mu\nu} G^{\prime
\alpha \mu\nu} 
\right]+\frac{\theta}{32\pi^2}\left[\sum_{a=1}^{8} G^a_{\mu\nu}
\widetilde{G}^{a\mu\nu}
+ \sum_{\alpha =1}^{N^{\prime 2}-1} G^{\prime \alpha}_{\mu\nu}
\widetilde{G}^{\prime \alpha \mu\nu} 
\right]~,
\label{4}
\end{equation}
where $G^a_{\mu\nu}$ and $G^{\prime \alpha}_{\mu\nu}$ denote the field
strength tensors of the ${\rm SU}(3)_c$ and ${\rm SU}(N')$ gauge
theories, respectively; $a =1,\dots 8$ is the ${\rm SU}(3)_c$ color
index while $\alpha = 1, \dots N^{\prime 2}-1$ is the ${\rm SU}(N')$
hidden color index. Consequently, at the scale $M_U$, the gauge couplings 
$g_s^{},g_s'$ and the theta parameters $\theta_s^{}, \theta_s'$ of the two gauge groups satisfy
\beq
g=g_s^{}= g'_s\,,\qquad \theta=\theta_s^{} = \theta^\prime_s\,.
\label{UVcond}
\eeq
In order to generate axion couplings  at a lower scale
compatible with assuming that the PQ symmetry is broken at
$10\,\text{TeV}$, we require that the strong coupling scale $\Lambda'$
of the hidden gauge group satisfy $1 \,\text{TeV} \lesssim \Lambda'
\lesssim 10\,\text{TeV}$. This requirement gives a strong constraint on
the numbers of hidden colors $N^\prime$ and hidden quark flavors
$n_F^\prime$. To see this, we first note that at the one-loop level the
strong coupling constant at the scale $M_U$ is given by
\begin{equation}
 \frac{1}{\alpha_s (M_U)} = \frac{1}{\alpha_s(m_Z)}
-\frac{b}{2\pi} \ln \left(\frac{M_U}{m_Z}\right) ~,
\end{equation}
where $\alpha_s \equiv g_s^2/(4\pi)$, $m_Z$ is the $Z$-boson mass,
$\alpha_s (m_Z) = 0.1185(6)$ \cite{Agashe:2014kda},\footnote{The
uncertainty in the input value of $\alpha_s (m_Z)$ causes less than
10\% errors for the resultant values of $M_U$ given in
Table~\ref{tab:muv}. } and $b = -7 +
\frac{2}{3}n_F^\prime$ with six quark flavors assumed. As we will see 
in more detail in Sec.~\ref{sec:axioncouplingsandmass},
the number of extra quark flavors is equal to that of the hidden quark
flavor $n_F^\prime$ in our setup since they originate from fundamental
representations of SU($3+N'$). On the other hand, the hidden coupling 
at $M_U$ is given by
\begin{equation}
  \frac{1}{\alpha_s^\prime (M_U)} = 
-\frac{b^\prime}{2\pi} \ln \left(\frac{M_U}{\Lambda^\prime}\right) ~,
\end{equation}
with $b^\prime = -\frac{11}{3}N^\prime
+\frac{2}{3}n_F^\prime$.\footnote{Strictly speaking, the coefficients
$b$ and $b^\prime$ should be modified below each extra-quark mass
threshold. However, since the extra quark masses (1--10~TeV) are not far
from the electroweak scale, we expect that one-step matching adopted
here does not cause significant uncertainty in this estimation.  } 
Here we assume that there are no mirror Standard Model quarks and
leptons at low energies. By requiring $\alpha_s (M_U) =
\alpha_s^\prime (M_U)$, we can express $M_U$ as a function
of $n_F^\prime$, $N^\prime$, and $\Lambda^\prime$.

\begin{table}[t!]
 \begin{center}
\caption{The values of $M_U$ (in GeV) for various $N^\prime$,
  $n_F^\prime$, and $\Lambda^\prime$.  }
\label{tab:muv}
\vspace{5pt}
\begin{tabular}{c|c|ll}
\hline
\hline
$N^\prime$ & $n_F^\prime$ & $\Lambda^\prime =1$~TeV & $\Lambda^\prime
 =10$~TeV \\
\hline
    & $1$ & $2.5\times 10^{10}$ & $9.7\times 10^{12}$ \\
    & $2$ & $1.7\times 10^{10}$ & $4.4\times 10^{12}$ \\
$3$ & $3$ & $1.1\times 10^{10}$ & $2.0\times 10^{12}$ \\
    & $4$ & $7.6\times 10^{9}$ & $9.3\times 10^{11}$ \\
    & $5$ & $5.1\times 10^{9}$ & $4.2\times 10^{11}$ \\
\hline
    & $1$ & $7.3\times 10^{6}$ & $4.9\times 10^{8}$ \\
    & $2$ & $5.9\times 10^{6}$ & $3.2\times 10^{8}$ \\
$4$ & $3$ & $4.8\times 10^{6}$ & $2.2\times 10^{8}$ \\
    & $4$ & $3.9\times 10^{6}$ & $1.4\times 10^{8}$ \\
    & $5$ & $3.2\times 10^{6}$ & $9.5\times 10^{7}$ \\
\hline
\hline
\end{tabular}
 \end{center}
\end{table}

In Table~\ref{tab:muv}, we summarize the values of $M_U$ (in GeV)
for various $N^\prime$, $n_F^\prime$, and $\Lambda^\prime$. It turns out
that the $N^\prime = 2$ cases do not yield any reasonable value for
$M_U$. For a larger $N^\prime$, we obtain a lower $M_U$. From this table, 
we find that this setup accommodates multi-flavors
for extra quarks while keeping $M_U$ sufficiently high. The more
vector-like quarks we add to the theory, the larger the beta function
of the hidden strong interaction becomes, which results in a smaller
coupling constant at low energies. On the other hand, these extra quarks
make the strong coupling constant larger at high scales, and thus the
unified coupling $g$ also becomes large. As these two effects compensate
each other, the resultant $M_U$ is rather insensitive to the
number of extra quarks. This feature is actually desirable for the explanation 
of the 750~GeV diphoton anomaly in our model, as we discuss in Sec.~\ref{sec:750}.

\begin{figure}[t]
\centering
\includegraphics[clip, width = 0.6 \textwidth]{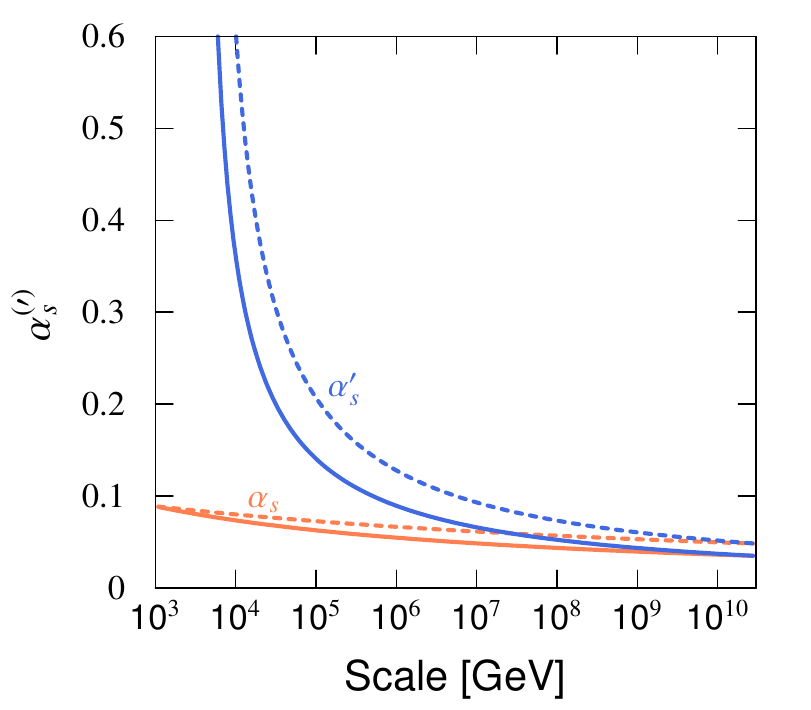}
\caption{The running of $\alpha_s$ and $\alpha_s^\prime$, where 
 $N^\prime = 3$ and $M_U = 3\times 10^{10}$~GeV. The solid and
 dashed lines correspond to the cases of $n_F^\prime =1$ and 5, respectively. } 
\label{fig:rge}
\end{figure}

In Fig.~\ref{fig:rge}, we show the running of $\alpha_s$ and
$\alpha_s^\prime$ with orange and blue lines for representative values, 
$N^\prime = 3$ and $M_U =
3\times 10^{10}$~GeV. The solid and dashed lines correspond to the cases of 
$n_F^\prime =1$ and 5, respectively. Here, we have used the two-loop renormalization 
group equations, and neglected threshold corrections at $M_U$. The masses of
the vector-like quarks are set to be 1~TeV. As can be seen,
$\Lambda^\prime$ is less sensitive to $n_F^\prime$, which allows us to
introduce a number of vector-like quarks at low energies. We note in
passing that our model does not suffer from a domain wall problem
\cite{Zeldovich:1974uw, Sikivie:1982qv} even though $n_F^\prime \geq
2$. As we will see in Sec.~\ref{sec:gravpq}, we can introduce the
PQ-symmetry violating Planck-suppressed operators without spoiling the
PQ mechanism. These operators explicitly break a discrete symmetry, and
thus destabilize domain walls.

\subsection{Axion couplings and mass}
\label{sec:axioncouplingsandmass}

We will assume that there are new Dirac quarks, $\Psi$ in the
fundamental representation of the unified color group ${\rm
SU}(3+N')$. After this group is spontaneously broken at the scale
$M_U$, these quarks split into a fundamental representation of
${\rm SU}(3)_c$, denoted $\psi$, and a fundamental
representation of the hidden color group ${\rm SU}(N')$, denoted
$\psi'$. In addition we assume that there is a complex scalar field
$\Phi$ that couples to the new Dirac fermions. As in the KSVZ model we
assume that these fields are charged under a Peccei--Quinn U(1) global
symmetry,  
\beq
     \Psi \rightarrow e^{i q_\Psi \alpha\gamma_5} \Psi\,, \qquad \Phi \rightarrow 
     e^{i q_\Phi \alpha} \Phi\,,
     \label{fermiontrans}
\eeq
where $\alpha$ is an arbitrary parameter and $q_{\Psi,\Phi}$ are the PQ
charges. We will assume $q_\Phi=1$ and $q_\Psi=\frac{1}{2}$ for simplicity. This symmetry forbids a Dirac mass term but allows the Yukawa couplings
\beqn
&&\Delta {\cal L} = h_{ij}\Phi \bar{\Psi}_{Ri} \Psi_{Lj} +{\rm h.c.}
\longrightarrow h_{ij}\Phi \left( \bar \psi_{Ri}
\psi_{Lj}+ \bar{\psi}^\prime_{Ri} \psi^{\prime }_{Lj} \right)+{\rm
h.c.}\,,
\label{eq:vecqyuk}
\eeqn
where $h_{ij}$ are dimensionless couplings and $i, j = 1,\dots
n_F^\prime$ denotes the flavor index. As one can see, the number of
extra quarks is equal to that of extra hidden quarks. The spontaneous
breaking of the PQ symmetry then occurs when the scalar field obtains a
VEV, which is parametrized as 
\beq
       \Phi= \frac{1}{\sqrt{2}} ( f_a + \rho) e^{i \frac{a}{f_a}}\,,
\eeq
where $f_a$ is the PQ breaking scale, $\rho$ is the radial mode and $a$ is the axion field. 
The radial mode obtains a mass of order $\sqrt{\lambda_\Phi} f_a$, where $\lambda_\Phi$ 
is the quartic coupling in the scalar potential. The PQ current becomes
\begin{align}
j_\mu^{\rm PQ} &= i q_\Phi\left( \Phi^* \partial_\mu \Phi 
-\Phi\partial_\mu \Phi^*
\right)
+ q_\Psi \bar \psi \gamma_\mu \gamma_5 \psi + q_\Psi \bar\psi' 
\gamma_\mu\gamma_5\psi'\nonumber\,,\\
&\to -f_a \partial_\mu a + \frac{1}{2} \bar \psi \gamma_\mu \gamma_5 \psi +
 \frac{1}{2} \bar\psi' \gamma_\mu \gamma_5 \psi' \,.
\label{PQcurrent}
\end{align}
Under a PQ transformation the axion will shift as $a\rightarrow a+f_a \alpha$,
giving rise to an anomalous term that matches the axial anomaly from (\ref{fermiontrans}).
Since the axion couples to the divergence of the PQ current, we see from Eq.~(\ref{PQcurrent}) 
that the axion couples to the new quarks $\psi,\psi'$, which obtain a mass of order $m_\Psi \sim h f_a$ 
after the PQ symmetry is broken. 

At low scales, these heavy quarks are integrated out (assuming $m_\Psi \gtrsim \Lambda'$) and generate a coupling of the axion field (and the radial field $\rho$) to the QCD gluons, the hidden sector gluons, and possibly photons (provided the heavy fermions also carry hypercharge). In particular,
\beq
{\cal L}_a= \frac{1}{32\pi^2} \left(\frac{a}{f_a}+\theta\right) \,G_{\mu\nu}^a\, \widetilde{G}^{a\mu\nu}\,,\qquad
{\cal L}_{a '} = \frac{1}{32\pi^2}
\left(\frac{a}{f_a}+\theta\right) \,{G}_{\mu\nu}^{\prime \alpha}\,
\widetilde{G}^{\prime \alpha\mu\nu}\,,
\label{effL}
\eeq
where we have used (\ref{UVcond}) and $\theta$ nonrenormalization. Note that the triangle graphs which generate (\ref{effL}) are saturated at virtual momenta $m_\Psi\sim h \langle \Phi \rangle\sim h f_a$.

The axion mass-squared is determined by the 
two-point function
\begin{align}
 &i\int d^4x \left\langle
F_{\mu\nu}^A\, \widetilde{F}^{A\mu\nu}(x) \,,
F_{\rho\sigma}^B\, \widetilde{F}^{B\rho\sigma}(0)
\right\rangle
\nonumber\\
\to&~
i\int d^4x \left\langle
G_{\mu\nu}^a\, \widetilde{G}^{a\mu\nu}(x) \,,
G_{\rho\sigma}^b\, \widetilde{G}^{b\rho\sigma}(0)
\right\rangle
+ i
\int d^4x \left\langle
G_{\mu\nu}^{\prime \alpha}\, \widetilde{G}^{\prime \alpha \mu\nu}(x) \,,
G_{\rho\sigma}^{\prime\beta}\, \widetilde{G}^{\prime \beta\rho\sigma}(0)
\right\rangle\,,
\end{align}
where the latter correlation function is saturated in the IR and reduces to $\sim
\frac 18\Lambda_c^4+ \Lambda^{\prime 4}$.
Since we deal with a single combination $a+\theta$, 
the axion Lagrangian takes the form
\beq
{\cal L}_a = \frac 12 \partial_\mu a\, \partial^\mu a - \frac 12 {\cal T}
\left(\frac{a}{f_a} +\theta \right)^2 ~,
\eeq
and thus the $\theta$-term is eliminated in the vacuum. 
Here, ${\cal T}$ is the sum of the topological susceptibilities for the
two Yang--Mills theories, QCD and the hidden color group. It is given by 
\beq
{\cal T} = \frac 18 \Lambda_c^4 +  \left(\Lambda'\right)^4\,;
\eeq
see Eq.~(\ref{2}) for the first term and the subsequent discussion for
the second. A dual interpretation of this mechanism is given in
Appendix~\ref{sec:dualpicture}. 

The axion mass relation then becomes
\beq
f_a^2 \, m_a^2 = {\cal T} \sim \frac 18\Lambda_c^4  + \left(\Lambda'\right)^4\,.
\label{axionmass}
\eeq
The second term on the right-hand side of (\ref{axionmass}) can be
arbitrarily large compared to the first term from QCD and, therefore, can
give the dominant contribution to the axion mass. This destroys the
standard KSVZ relation between $m_a$ and $\Lambda_c$ allowing for much larger
values of the axion mass. For example, for $\Lambda' \sim \,\text{1
TeV}$ and $f_a\sim 10\, \text{TeV}$, the axion mass $m_a$ can be as
large as ${\cal O}(100)$ GeV!
This then invalidates the standard axion limits from astrophysics.

\subsection{Unified symmetry breaking effects}

After the spontaneous breaking of ${\rm SU}(3+N')$, there could be
possible sources of CP violation that spoil the equation $\theta_s = \theta'_s$ 
at low energies, since the physical theta parameter is given by
\beq
     \bar\theta = \theta + \text{arg}(\text{det}{\cal M})\,,
\eeq
where ${\cal M}$ is a complex mass matrix for quarks. These include threshold 
effects and renormalization group effects caused by visible and hidden quarks, 
and higher-dimensional operators that contain the ${\rm SU}(3+N')$ breaking field $\Sigma$.  

Firstly, we consider the effects of vector-like quarks on the vacuum
angles. Above $M_U$, the vector-like quarks form the fundamental
representation of SU($3+N^\prime$), and they have Yukawa couplings with
the scalar field $\Phi$ as in Eq.~\eqref{eq:vecqyuk}. Below $M_U$, the 
Yukawa interaction splits into two parts as shown in the
right-hand side of Eq.~\eqref{eq:vecqyuk}, but the coefficients of the
two parts, $h_{ij}$, are identical. For this reason, after $\Phi$
develops a VEV, the resultant mass matrices for $\psi$ and $\psi^\prime$
also become identical, $h f_a/\sqrt{2}$. Therefore, these mass terms
contribute to the $\theta$ angles with the same amount, ${\rm arg}\{{\rm
det} (hf_a/\sqrt{2})\}$, and do not spoil the relation $\theta_s =
\theta'_s$. 

Secondly, we consider the contribution of the Standard Model quarks,
$Q_L$, $u^c_R$, and $d^c_R$, and their SU$(3+N^\prime)$ partners, 
$Q^\prime_L$, $u^{\prime c}_R$, and $d^{\prime c}_R$,
respectively. These fields form fundamental representations of ${\rm
SU}(3+N^\prime)$, $\Psi_Q$, $\Psi_{\bar{u}}$, and $\Psi_{\bar
d}$.\footnote{This assumes that there is an anomaly-free UV completion,
where the local SU$(3+N^\prime)$ gauge anomalies cancel. This requires
extra UV states which can be decoupled at $M_U$ without affecting our
arguments.}  
The leptons are irrelevant for the present discussion and thus we will
neglect them in what follows.  As we will see, there are subtleties in
this case since the low-energy spectrum of our model does not contain
the partner quarks, and thus the SU$(3+N^\prime)$
symmetry is explicitly broken in this sector.

These fields have Yukawa interactions with the Standard Model Higgs
boson in order to reproduce the ordinary Standard Model Yukawa
couplings. In the ${\rm SU}(3+N^\prime)$ gauge theory, these Yukawa
interactions are written as  
\begin{equation}
 {\cal L}_{\rm Yukawa} = -\Psi_{Qi} ({\cal Y}_u)_{ij} \Psi_{\bar{u}j} H 
- H^\dagger \Psi_{Qi} ({\cal Y}_d)_{ij} \Psi_{\bar{d}j} +{\rm h.c.} ~,
\end{equation}
where $i,j = 1,2,3$ is the generation index, 
${\cal Y}_u$ and ${\cal Y}_d$ are $3\times 3$ matrices, and $H$ is the Standard
Model Higgs field. Since the values of the theta terms are basis-dependent, we first
specify the basis for the following discussion. Of course, the
derived consequences do not depend on the choice of the basis. 

Using the possible field re-definitions, the Yukawa matrices can be transformed to the
following form:
\begin{equation}
 {\cal Y}_u = \text{diag}(y_u, y_c, y_t)~, ~~~~~~
 {\cal Y}_d = V_{\rm CKM}^* \cdot \text{diag}(y_d, y_s, y_b)\,,
\label{eq:yukawaform}
\end{equation}
where $V_{\rm CKM}$ is the ordinary CKM matrix. As discussed in
Sec.~\ref{sec:couptheta}, we have $\theta_s = \theta_s^\prime$ below the
SU($3+N^\prime$) symmetry breaking scale. On the other hand, the Yukawa
interactions lead to  
\begin{equation}
 {\cal L}_{\rm Yukawa} = -Q_L {\cal Y}_u u^c_R H -H^\dagger Q_L {\cal Y}_d d^c_R 
-Q_L^\prime {\cal Y}_u u^{\prime c}_R H -H^\dagger Q^\prime_L {\cal Y}_d d^{\prime
c}_R +{\rm h.c.} ~. 
\label{eq:SMHiggsYukawa}
\end{equation}
Now let us examine the physical $\theta$ terms of both sectors. In
the SU(3)$_c$ sector,
\begin{align}
 \bar{\theta} \equiv \theta_s + {\rm arg}({\rm det} {\cal Y}_u)
+ {\rm arg}({\rm det} {\cal Y}_d) 
= \theta_s - {\rm arg}({\rm det} V_{\rm CKM}) = \theta_s~,
\end{align}
where we have used $\det (V_{\rm CKM}) = 1$.

On the other hand, the physical vacuum angle in the SU($N^\prime$)
sector depends on the mass splitting mechanism for $Q^\prime$,
$u^\prime$, and $d^\prime$. If the mass splitting mechanism does not
introduce new CP phases, which can be naturally realized with, 
{\it e.g.}, a warped extra dimension compactified on an orbifold (see
Appendix~\ref{sec:decpart}), then again we have  $\bar{\theta}^\prime 
= \theta_s^\prime$.
Thus, we conclude that
\begin{equation}
 \overline{\theta} = \overline{\theta}^\prime ~,
 \label{eq:unif_theta}
\end{equation}
in the unified model, assuming that the SU($3+N^\prime$)-preserving CP violation 
is sufficiently sequestered from the symmetry breaking. Once this relation holds at
$M_U$, it is not spoiled at low energies since the physical theta
terms are invariant under renormalization group flow. 

Finally, we consider the effects of higher-dimensional CP-odd operators
including the SU($3+N^\prime$)-breaking field $\Sigma$, which are
expected to be induced at the Planck scale $M_P$ ({\it e.g.} by virtual black holes). 
Among them, the following dimension-five operator gives the dominant effect: 
\begin{equation}
 \frac{c}{M_P} {\rm Tr}(\Sigma F_{\mu\nu}\widetilde{F}^{\mu\nu}) ~,
 \label{Planckdim5}
\end{equation}
where $F_{\mu\nu} \equiv F^A_{\mu\nu}T^A$ with $T^A$ the generators of
SU($3+N^\prime$) and $c$ is a dimensionless constant. This operator
reduces to a theta term after $\Sigma$ gets a VEV (see Eq.~\eqref{sigvev}), 
and thus could spoil the relation $\theta_s = \theta'_s$. This, however, 
causes no problem if $|c\langle \Sigma \rangle| < 10^{-10} M_P\simeq 2\times 
10^8$~GeV. This can be naturally realized for
$N^\prime = 4$, as can be seen in Table~\ref{tab:muv}. For $N^\prime
=3$, the above limit gives $|c|\lesssim 10^{-2}$. Thus, we see that the theta
relation in Eq.~(\ref{UVcond}) can be well maintained in the IR, so that the
axion can cancel both theta terms. 

\subsection{Gravitational violations of PQ symmetry}
\label{sec:gravpq}

An immediate consequence of an electroweak scale axion is that
gravitational violations of the PQ global symmetry become naturally
suppressed~\cite{Berezhiani:2000gh}. Below the Planck scale, the
effective PQ-violating terms are described by the Planck-scale-suppressed 
higher-dimensional operators\footnote{Here we assume that the PQ
symmetry is broken only through higher-dimensional operators, though
renormalizable operators can also be present if, for instance, wormhole
effects are sizable \cite{Holman:1992us}. } 
\beq
         {\cal L} =  \frac{\kappa}{M_P^{2m+n-4}}|\Phi|^{2m}\Phi^{n} + {\rm h.c.}\,,
         \label{gravLag}
\eeq
where $\kappa$ is a dimensionless constant and $m,n$ are integers satisfying
$n\geq 1$ and $2m+n\geq 5$.
Such an operator induces an effective $\theta$-angle
\cite{Kamionkowski:1992mf, Holman:1992us, Barr:1992qq}
\begin{equation}
 \theta_{\rm eff} \sim |\kappa|\left(\frac{f_a}{m_a}\right)^2
  \left(\frac{f_a}{\sqrt{2}M_P}\right)^{2m+n-4} ~,
\end{equation}
where we have omitted an ${\cal O}(1)$ factor for brevity. In particular,
dimension-five operators ($2m+n=5$) generate an effective $\theta$-angle of 
\beq
       \theta_{\rm eff} \sim 10^{-12}\times|\kappa|\cdot \left(
       \frac{f_a}{10~\text{TeV}} \right)^3 
\left(\frac{750~\text{GeV}}{m_a}\right)^2~.
\eeq
This value is sufficiently suppressed for the electroweak scale axion
that it does not spoil the axion mechanism. This contrasts with the
usual invisible axion models where since $f_a\gtrsim 10^9$ GeV,
gravitational PQ-symmetry violating terms to very high order ($n\gtrsim
10$) must be suppressed \cite{Kamionkowski:1992mf, Holman:1992us,
Barr:1992qq}.

However, in the presence of extra Higgs fields which develop
large VEVs, such as the SU($3+N^\prime$) breaking Higgs field $\Sigma$,
there could be other PQ-violating operators like $|\Sigma|^{2m} \Phi^n
/M_P^{2m+n-4}$, which may spoil the PQ mechanism. We thus assume that
such operators are sufficiently suppressed. Note however, that the
SU($3+N^\prime$) gauge group can be broken without the $\Sigma$ field if
we consider unification with an extra dimension compactified on an
orbifold. In this case, the above problem can be avoided.

\section{Phenomenological Consequences}

\subsection{The electroweak axion}
\label{sec:EWaxion}

Intriguingly, in our model,  the value of the axion mass can be in the several hundred
GeV range for a confinement scale, $\Lambda' \sim$ TeV and a PQ breaking
scale, $f_a\sim 10$ TeV. This axion is therefore quite ``visible'' and can be searched for
in collider experiments. As shown in Table~\ref{tab:muv}, such a confinement 
scale is obtained with $N^\prime = 3,4,\dots$. For concreteness, we choose $N'=3$ 
and assume that the QCD color group is embedded into ${\rm SU}(6)$ in what follows. 
Including the electroweak sector, the complete gauge group is 
${\rm SU}(6)\times {\rm SU}(2)_L\times {\rm U}(1)_Y$. 

We consider a set of vector-like quarks, $\Psi$ transforming in the
${\bf 6} \oplus {\bf{\bar 6}}$ of SU(6). They are supposed to be singlets
under the SU(2)$_L$ gauge interaction. After SU(6) is broken to 
${\rm SU}(3)_c\times {\rm SU}(3^\prime)$ 
we obtain a pair of QCD Dirac fermions, $\psi$ transforming as 
$({\bf 3},{\bf 1})_{Y}\oplus ({\bf\bar 3},{\bf 1})_{-Y}$, and a pair, $\psi^\prime$ 
transforming as $({\bf 1},{\bf 3})_{Y'}\oplus ({\bf 1},{\bf\bar 3})_{-Y'}$ of the 
hidden color group, where $Y$ and $Y^\prime$ are the Standard Model
hypercharges.\footnote{Note that even though $Y=Y'$ when the U(1) subgroup
of SU(6) is broken in the way described after (\ref{sigvev}), we allow the more general possibility 
that $Y\neq Y'$ which can occur when a linear combination of the U(1) subgroup of SU(6) 
and an additional U(1) is broken to give the usual U(1)$_Y$ hypercharge below the unification scale. 
For example, this occurs when the scalar of the three-index antisymmetric tensor is charged under the
additional U(1).}
When integrated out, these fermions generate the effective axion couplings to 
gluons and photons:   
\begin{equation}
{\cal L}_a= n_F^\prime\frac{\alpha_s}{8\pi} \frac{a}{f_a} G_{\mu\nu}^a\,
 \widetilde{G}^{a\mu\nu}+ 6n_F^\prime(Y^2 + Y^{\prime 2})
\frac{\alpha_Y}{8\pi}
 \frac{a}{f_a} B_{\mu\nu}\, \widetilde{B}^{\mu\nu}\,,
 \label{axionLag}
\end{equation}
where $\alpha_s \equiv g_s^2/(4\pi)$, $\alpha_Y \equiv g_Y^2/(4\pi)$ with $g_Y$ 
the coupling constant of the U(1)$_Y$ gauge interaction, and $B_{\mu\nu}$ the 
hypercharge field strength tensor. Note that we have moved to the basis where the 
gauge fields are canonically normalized. Only $\psi$ contributes to the first term on 
the right-hand side of (\ref{axionLag}), while both $\psi$ and $\psi^\prime$ generate 
the second term. We also note that in the electroweak symmetry breaking basis, 
\begin{equation}
 \alpha_Y B_{\mu\nu} \widetilde{B}^{\mu\nu}
= \alpha_{\rm EM} \left[F_{\mu\nu} \widetilde{F}^{\mu\nu}
-2 \tan\theta_W F_{\mu\nu} \widetilde{Z}^{\mu\nu}
+\tan^2\theta_W Z_{\mu\nu} \widetilde{Z}^{\mu\nu}
\right]~,
\end{equation}
where $\alpha_{\rm EM}$ denotes the fine-structure constant, $\theta_W$
is the weak-mixing angle, and $F_{\mu\nu}, Z_{\mu\nu}$ are the field
strength tensors for the photon and $Z$-boson, respectively.  

An electroweak scale axion, $a$ is produced at the LHC via the gluon fusion
process. The production cross section is given by
\begin{equation}
 \sigma (pp \to a) = \frac{k_g}{m_a s} C_{gg} \Gamma(a\to gg)~,
 \label{axionproduction}
\end{equation}
where $m_a$ is the the axion mass, $\sqrt{s}$ is the center-of-mass energy 
of the $pp$ collision, and $C_{gg}$ is the gluon luminosity factor defined by
\begin{equation}
 C_{gg} =\frac{\pi^2}{8}\int dx_1 dx_2 \delta(x_1 x_2 -m_a^2/s)g(x_1) g(x_2)~,
\end{equation}
with $g(x)$ the gluon parton distribution function (PDF). The so-called $k$-factor, $k_g$ is a 
multiplicative factor that parametrizes higher-order QCD corrections. The partial decay 
width of the axion into a pair of gluons, $\Gamma(a \to gg)$ is given by
\begin{equation}
 \Gamma (a \to gg) = \frac{\alpha_s^2}{32\pi^3} \frac{n_F^{\prime 2} m_a^3}{
  f_a^2} ~.
\end{equation}
Notice that $\Gamma (a \to gg)$, and thus the production cross section is inversely
proportional to the square of $f_a/n_F^{\prime}$. 

Once produced, the axion decays into $gg$, $\gamma \gamma$, $Z \gamma$,
or $ZZ$. The partial decay widths of $\gamma \gamma$, $Z \gamma$, and
$ZZ$ are 
\begin{align}
  \Gamma (a \to \gamma \gamma) &= \frac{9 \alpha_{\rm EM}^2}{64\pi^3} 
  (Y^2+Y^{\prime 2})^2  \frac{n_F^{\prime 2} m_a^3}{f_a^2}~, \\
\Gamma (a \to Z \gamma) &\simeq 2\tan^2\theta_W \Gamma(a\to
 \gamma\gamma) ~,  \\
\Gamma (a \to Z Z) &\simeq \tan^4\theta_W \Gamma(a\to
 \gamma\gamma) ~,
\end{align}
respectively. Note that these decay widths are related to each other via
$\tan \theta_W\simeq 0.55$. In particular, the $ZZ$ decay mode is significantly
suppressed by a factor of $\tan^4 \theta_W$ compared to the diphoton
decay channel. 

In our minimal model, we have assumed that the electroweak axion has no coupling to
$W$ bosons. However $W$-boson couplings can be generated by introducing
vector-like fermions charged under SU(2)$_L$. Furthermore, since the Standard Model
quarks and leptons are not charged under the PQ symmetry, as in the
original KSVZ model, there are no tree-level axion couplings to Standard
Model fermions. These couplings are instead induced at higher-loop level compared 
with the photon and gluon couplings, and thus negligible in the present analysis. 

Besides the axion, the model also predicts colored vector-like fermions
at a mass scale $\sim h f_a$, where $h$ is a Yukawa coupling. Depending
on the value of $h$, these fermions may be near the TeV scale. Furthermore, 
the radial scalar mode, $\rho$ will obtain a mass of order $\sqrt{\lambda_\Phi} f_a$,
where $\lambda_\Phi$ is the quartic coupling of the complex  scalar, $\Phi$ potential. 
Thus our model has quite minimal predictions, which can be probed at Run-II of
the LHC.

\subsection{The 750 GeV diphoton resonance}
\label{sec:750}

Recently, the ATLAS \cite{ATLASdiphoton, ATLASdiphoton2} and CMS
\cite{CMS:2015dxe, CMS:2016owr}
collaborations announced an excess of events around 750~GeV in
the diphoton resonance searches at the 13~TeV LHC run. These excesses
can be explained if the production cross section of the 750~GeV
resonance times its decay branching fraction to diphotons is 5--10~fb. 
After the announcement, many possible explanations have been
proposed~\cite{Harigaya:2015ezk,
Mambrini:2015wyu, Backovic:2015fnp, Angelescu:2015uiz, Nakai:2015ptz,
Knapen:2015dap, Buttazzo:2015txu, Pilaftsis:2015ycr,
Franceschini:2015kwy, DiChiara:2015vdm, Higaki:2015jag,
McDermott:2015sck, Ellis:2015oso, Low:2015qep, Bellazzini:2015nxw,
Gupta:2015zzs, Petersson:2015mkr, Molinaro:2015cwg, Staub:2016dxq,
Arcadi:2016dbl}.  

Obviously the electroweak axion in our model can be a candidate for the 750 GeV
resonance.\footnote{For other models which consider the interplay between
the 750 GeV resonance and a solution to the strong CP problem (or
axion), see Refs.~\cite{Pilaftsis:2015ycr, Higaki:2015jag, Cao:2015xjz,
Kim:2015xyn, Ben-Dayan:2016gxw, Barrie:2016ntq, Aparicio:2016iwr,
Kobakhidze:2016wmv, Chiang:2016eav, Higaki:2016yqk}. } 
Identifying the visible axion with the 750 GeV resonance requires that  
\beq
     m_a \sim \frac{(\Lambda')^2}{f_a} \sim 750\,\text{GeV}\,,
\eeq
or equivalently 
\begin{equation}
 \Lambda^\prime \sim \left(\frac{f_a}{1~{\rm TeV}}\right)^{\frac{1}{2}}
  \times 870~{\rm GeV} ~.
\end{equation}
The 750~GeV axion is produced at the LHC via the gluon fusion
process.\footnote{Photo-production is negligible unless the hypercharges
$Y$ and $Y^\prime$ are very large.} The production cross section can be calculated using
(\ref{axionproduction}) where the numerical value of $C_{gg}$ is evaluated using the 
{\tt MSTW2008NLO} PDF data set \cite{Martin:2009iq} in Ref.~\cite{Franceschini:2015kwy} 
as $C_{gg} \simeq 2137$ (174) for $\sqrt{s} = 13$~TeV (8~TeV), and the $k$-factor is taken 
to be $k_g \simeq 2$ \cite{Baglio:2010ae}.  

\begin{figure}[t]
\begin{center}
\subfigure[Production cross section]
 {\includegraphics[clip, width = 0.48 \textwidth]{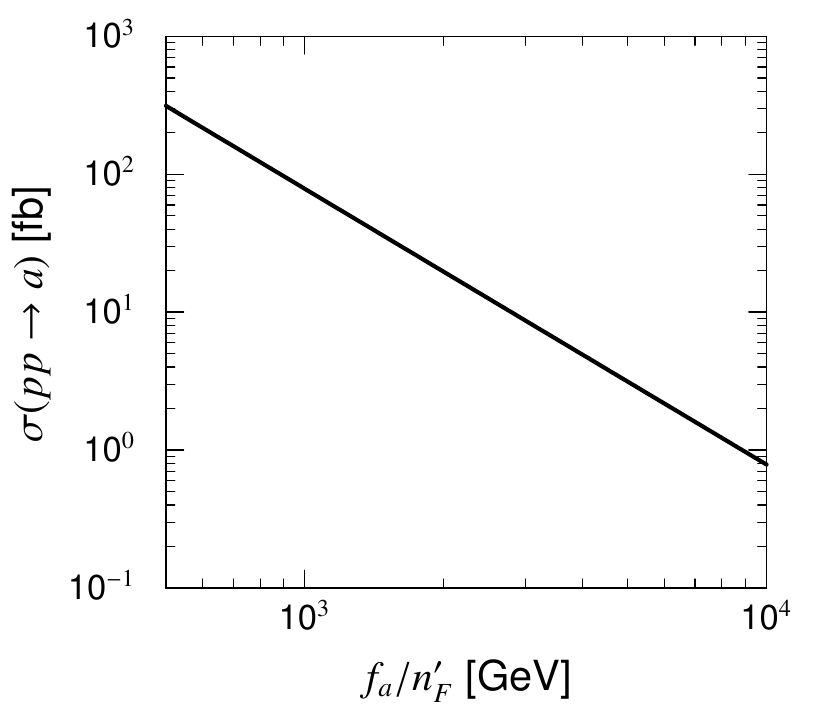}
 \label{fig:prodcross}}
\subfigure[Branching ratios]
 {\includegraphics[clip, width = 0.48 \textwidth]{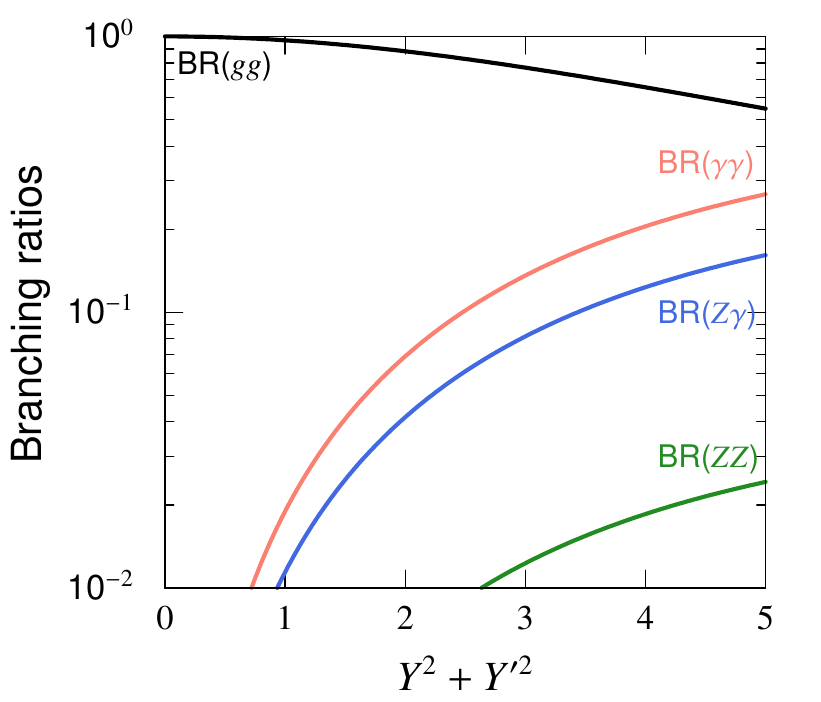}
 \label{fig:br}}
\caption{The production cross section and branching ratios of the axion
 $a$, assuming $m_a = 750$~GeV. (a) Production cross section $\sigma
 (pp \to a)$ as a function of $f_a/n_F^\prime$. (b) Branching ratios of
 $a$ as functions of $Y^2 + Y^{\prime 2}$. }
\label{fig:prbr}
\end{center}
\end{figure}

In Fig.~\ref{fig:prodcross}, we show the axion production cross section 
as a function of $f_a/n_F^\prime$ assuming $m_a = 750$~GeV.
Given that the observed diphoton rate implies a signal cross
section of 5--10~fb, we see that the 750~GeV axion can explain the
diphoton excess if $f_a/n_F^\prime \sim 1$~TeV and the branching
fraction of the axion into diphotons is sizable.

In Fig.~\ref{fig:br}, we show the axion branching ratios as functions
of $Y^2 + Y^{\prime 2}$ where black, red, blue, and green lines (from top
to bottom) represent the branching fractions into dijet (a pair of gluons), 
diphoton, $Z\gamma$, and $ZZ$ channels, respectively. From this figure,
we find that a sizable rate into diphotons can be easily realized in
our model; for instance, $Y = Y^\prime = 1$ gives ${\rm BR}(a \to
\gamma \gamma) \simeq 7$\%. Note, however that if hypercharges $Y$ and
$Y^\prime$ are very large (or have (unusual) irrational values), stable 
charged particles (such as the lightest baryon composed of three $\psi^\prime$s) 
may appear, which are cosmologically problematic. These charged particles can 
decay into Standard Model particles via interactions described by effective 
higher-dimensional operators. If $Y$ and $Y^\prime$ are very large, such 
operators containing Standard Model fields must have correspondingly large 
dimensions since the hypercharges of the Standard Model particles are $\leq 1$. 
Therefore, in order for the charged particles to have a sufficiently
short lifetime, there must be a new scale below the unification scale,
$M_U$ at which these operators can be generated. Instead, the
fact that $Y, Y^\prime \sim 1$ gives rise to a sizable diphoton
branching ratio suggests that there exists a simple UV model with
operators generated at or above the UV scale which does not have charged
stable particles and can explain the 750~GeV diphoton events. 

For example, consider a set of vector-like quarks $\psi^{(\prime)}_u$ and
$\psi^{(\prime)}_d$ which have hypercharges $Y^{(\prime)} =\frac{2}{3}$
and $-\frac{1}{3}$, respectively. If $\psi^{(\prime)}_u$ is
heavier than $\psi^{(\prime)}_d$, the lightest baryon is
composed of one $\psi^{(\prime)}_u$ and two $\psi^{(\prime)}_d$s, which
is electrically neutral and thus can be a dark matter candidate, assuming it is
stable. The heavier charged baryon, which is composed of two
$\psi^{(\prime)}_u$s and one $\psi^{(\prime)}_d$, can decay if
we introduce, for instance, a charged scalar $\phi^+$ with a PQ charge
$+1$. This charged scalar can have a Yukawa coupling
$\bar{\psi}^{(\prime)}_{uR} \psi^{(\prime)}_{dL} \phi^+$ as well as a coupling to
the Standard Model sector via a dimension-five operator like $\phi^+
\Phi^* \bar{u}_Rd_R$, which can be induced at $M_U$ via a
trilinear coupling $\phi^+ \varphi^- \Phi^*$ and a Yukawa coupling 
$\varphi^+ \bar{u}_Rd_R$ where $\varphi^\pm$ are charged scalars with zero PQ
charge and mass of ${\cal O}(M_U)$. The introduction
of these fields and interactions does not spoil the relation $\overline{\theta} =
\overline{\theta}^\prime$ as they do not induce mass terms for
fermions.\footnote{Note that the unprimed fields will form visible baryons as well
as R-hadron-like states with Standard Model quarks. These heavy bound 
states ($\gtrsim$~TeV) can be made to decay promptly, and could eventually 
be detected at a future collider.}
An alternative possibility is to embed our model into an SU(2)$_R$ gauge theory
above $M_U$ by putting $\psi^{(\prime)}_u$ and $\psi^{(\prime)}_d$ into 
a fundamental representation of SU(2)$_R$ with the Standard Model fields 
also embedded into SU(2)$_R$ representations in the usual manner. In this case, 
$\psi^{(\prime)}_u$ can decay into $\psi^{(\prime)}_d$ plus the Standard Model 
particles via the exchange of a SU(2)$_R$ gauge boson with an 
${\cal O}(M_U)$ mass. Thus, we see that there are various possible ways to 
incorporate dark matter in a UV completion.

\begin{figure}[t]
\centering
\includegraphics[clip, width = 0.7 \textwidth]{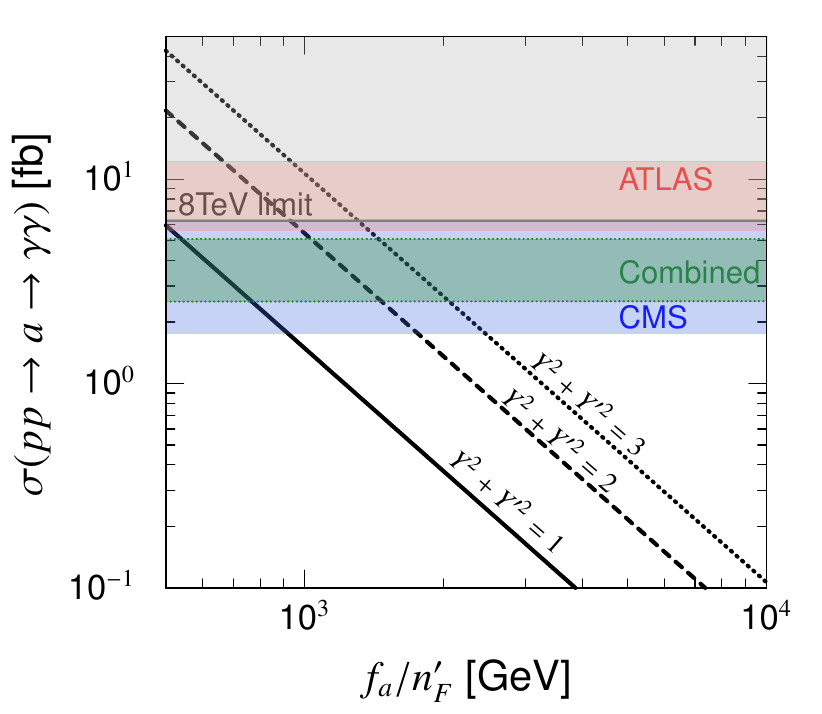}
\caption{Cross sections of the diphoton resonance events as functions of
 $f_a/n_F^\prime$, where the black solid, dashed, and dotted lines show
the cases of $Y^2+Y^{\prime 2} =1$, 2, and 3, respectively. The red
(blue) shaded area reproduces the number of events observed by the ATLAS
\cite{ATLASdiphoton} (CMS \cite{CMS:2015dxe}) collaboration. The
gray-shaded region is disfavored by the 8~TeV results \cite{Aad:2014ioa,
Khachatryan:2015qba}. The green shaded region corresponds to the
best fit cross section obtained in Ref.~\cite{Buckley:2016mbr}.} 
\label{fig:cross}
\end{figure}

Next we evaluate the cross sections of the diphoton resonance events
predicted in this model. We plot them as functions of $f_a/n_F^\prime$ in
Fig.~\ref{fig:cross}. This figure shows that the diphoton excess can be 
explained if $f_a/n_F^\prime \sim 1$~TeV and the hypercharges are ${\cal O}(1)$. 
For example, when $Y=Y^\prime =1$, the best-fit cross section is achieved with 
$f_a/n_F^\prime =1$--1.5~TeV. This corresponds to a total width $\Gamma_{\rm tot}= 3$--6~MeV
and predicts the $Z\gamma$ cross section $\simeq 1.5$--3~fb and the dijet cross section $\simeq 32$--65~fb.
Notice that $n_F^\prime \gg 1$ is possible as discussed in
Sec.~\ref{sec:enqcd}. Thus, $f_a$ can be as large as 10~TeV if one
introduces a sufficient number of extra vector-like fermions. This means
that vector-like quarks and the radial scalar mode $\rho$ will have
masses in the multi-TeV range depending on the values of the Yukawa
coupling $h$ and quartic coupling $\lambda_\Phi$, respectively. If the
vector-like quarks are heavier than the CP-even scalar $\rho$, then it
can only decay to axion pairs, otherwise the $\rho$ will decay into
(possibly) long-lived vector-like quarks as well. If a glueball made of
the SU($N^\prime$) gluons has a mass smaller than half the $\rho$ mass,
then $\rho$ can also decay into a pair of hidden glueballs at the one-loop level.

Visible vector-like quarks can also be directly produced via strong interactions, 
and thus can be a good target at the LHC. They are observed as long-lived 
heavy hadrons, which may have an exotic electromagnetic charge depending on
their hypercharge. Hidden vector-like quarks\footnote{Phenomenological
aspects of such particles were first discussed by Okun in
Refs.~\cite{Okun:1980kw, Okun:1980mu}, where these particles were dubbed
``theta-leptons''. More recently their collider phenomenology was discussed in
Ref.~\cite{Kang:2008ea}, where they are referred to as ``quirks'' (see
Refs.~\cite{Foot:1989zk, Carlson:1991zn} for earlier work). 
Quirks have also recently been discussed in connection with the 750~GeV
anomaly; see, for instance, Refs.~\cite{Curtin:2015jcv, Agrawal:2015dbf,
Craig:2015lra}. } are, on the other hand, produced only through the U(1)$_Y$ 
gauge interaction, and thus their production cross sections are rather small. 
Nevertheless, they may be probed at Run-II of the LHC since they yield quite 
distinct signatures. As soon as hidden vector-like quarks are pair-produced, they annihilate
promptly, and can be observed as dilepton, dijet, and diphoton resonances. They can 
also annihilate into hidden glueballs leading to a similar phenomenology as that 
considered in Ref.~\cite{Juknevich:2009ji}.

\section{Conclusion}

In this paper, we have generalized the existing axion solution to allow for the possibility of a  
much heavier, visible axion. This is done by enlarging the QCD color group, 
${\rm SU}(3)_c$ to be ${\rm SU}(3+N')$ which is then broken to ${\rm SU}(3)_c\times {\rm SU}(N')$ at a UV scale, generating equal theta terms for the two gauge groups. 
Moreover due to the unified structure, the CP-violating contributions from complex mass matrices are identical in the two sectors. This requires that the ${\rm SU}(3+N')$-preserving CP violation is sufficiently sequestered from 
symmetry-breaking effects and no new phases are introduced when the unified partners of the Standard Model quarks are decoupled. In addition to the Standard Model quarks, there are extra vector-like quarks charged under a global PQ symmetry. After the PQ symmetry is spontaneously broken at a scale $f_a$, the extra vector-like quarks can be integrated out, generating a dimension-five axion coupling to gluons and, possibly photons. The unified origin of the theta and Yukawa terms then guarantees that after nonperturbative effects generate an axion potential, the two theta parameters can both be cancelled by a single axion.

Since the quark matter content is different between the two sectors, the
${\rm SU}(N')$ group can confine at a scale, $\Lambda'$ much larger than in QCD. This then gives the dominant contribution to the axion mass, thereby untying the usual dependence between the axion mass $m_a$ and the QCD confinement scale $\Lambda_c$. This gives rise to a model 
more flexible than the KSVZ invisible axion with regards to accommodating experimental data. For example, if $\Lambda'\sim$~TeV and the PQ breaking scale $f_a\sim 10$~TeV, 
then the axion obtains an electroweak scale mass.  Thus, our model describes a ``visible'' axion which can be (or perhaps, already was) detected in experiments.

Although it is true that the construction we develop is more complicated and less elegant than the classical invisible axion it may open a window into a new corner of ``beyond the Standard Model'' physics. First of all, an electroweak axion is theoretically aesthetic because it helps to suppress gravitational violations of the global PQ symmetry. Secondly, it changes the pattern of expectation established from cosmology and astrophysics, completely opening up the axion ``window''. Finally, it is irresistible not to identify our visible axion as a candidate for the explanation of the 750 GeV diphoton peak at the LHC, assuming it survives with more experimental data. In the minimal model, it predicts decays to dijets, $Z\gamma$ and $ZZ$, as well as new states such as vector-like quarks and a CP-even scalar mode with masses in the multi-TeV scale. Otherwise, if the signal disappears, the electroweak axion can still be searched for in future experiments together with the vector-like quarks and the PQ scalar mode, in order to establish whether or not Nature prefers this more unified solution of the strong CP problem.

\section*{Acknowledgments}

M.S. is thankful to G. Dvali for a prolonged debate regarding
Refs.~\cite{Dvali:2005an, gia_argument}, and P. Sikivie for useful correspondence. Helpful
discussions with Z. Berezhiani, K. Howe, A. Pomarol, and A. Vainshtein 
are gratefully acknowledged. We also thank M. Asano, S. Dimopoulos, A. Hook, M. Ibe, G. Moore, and K. Tobioka 
for valuable comments. 
This work was supported by the DOE grant DE-SC0011842 at
the University of Minnesota. 

\section*{Appendix}
\appendix

\section{A possible UV description}
\label{sec:decpart}

\renewcommand{\theequation}{A.\arabic{equation}}
\setcounter{equation}{0}

Our low-energy model crucially depends on not introducing CP phases when the
unified partners of the Standard Model quarks are decoupled. A UV framework to address this 
issue is to consider a warped extra dimension compactified on a $Z_2$-orbifold, 
where the SU($3+N^\prime$) gauge fields as well as the Standard Model quarks 
and their partners propagate in a CP-preserving bulk (with the SU(2)$_L\times$U(1) symmetry 
implicitly assumed). The UV brane (identified with a scale near the Planck scale) is also assumed 
to be SU($3+N^\prime$) symmetric, but CP is not conserved. It provides the source of CP violation including 
terms like in (\ref{UVLag}) and (\ref{Planckdim5}), as well as in the Higgs Yukawa coupling (\ref{eq:SMHiggsYukawa}) to 
Standard Model quarks and their partners. Furthermore, the PQ-charged vector-like quarks $\Psi$ and the 
PQ scalar field $\Phi$, are confined to the UV brane with the SU($3+N^\prime$)-symmetric Yukawa coupling (\ref{eq:vecqyuk}). 

Boundary conditions are then chosen to break the bulk gauge symmetry 
to ${\rm SU}(3)\times {\rm SU}(N^\prime)$ on the IR brane (identified with the 
$M_U$ scale), so that only the ${\rm SU}(3)\times {\rm SU}(N^\prime)$ gauge fields 
and the Standard Model quarks have massless zero modes.  This is similar to orbifold 
grand-unified models where only the Standard Model gauge bosons and the electroweak
Higgs fields have massless zero modes \cite{Kawamura:1999nj, Kawamura:2000ev, 
Kawamura:2000ir, Altarelli:2001qj, Hall:2001pg}. We further assume that the IR brane 
preserves the CP symmetry so that the quark partner fields are projected out without 
introducing extra CP phases.\,\footnote{Note that on the IR brane the boundary gauge 
couplings can be different, but we assume that the bulk contribution dominates.} 
Thus, the SU($3+N^\prime$)-symmetric CP violation on the UV brane is ``shined'' onto 
the CP-preserving ${\rm SU}(3)\times {\rm SU}(N^\prime)$ IR brane, realizing the 
condition (\ref{eq:unif_theta}) at the scale $M_U$.

The warped dimension also admits a dual four-dimensional interpretation via the AdS/CFT 
correspondence. The source of CP violation is confined to an elementary sector containing
SU($3+N^\prime$) gauge fields, vector-like fermions $\Psi$ and the PQ complex scalar field 
$\Phi$. The SU($3+N^\prime$) elementary gauge fields weakly gauge the SU($3+N^\prime$) global 
symmetry of some (unknown) strong ``technicolor'' dynamics. The strong dynamics preserves CP 
(via possibly massless ``techniquarks'') and spontaneously breaks the global symmetry to 
${\rm SU}(3)\times {\rm SU}(N^\prime)$. The corresponding gauge fields remain massless and the 
Standard Model quark partners obtain a mass of order the confinement scale of the strong dynamics. 
The source of CP violation is again SU($3+N^\prime$)
symmetric, realizing the initial conditions at $M_U$ for our visible axion model.

\subsection{A field theory example of decoupling quarks}
The orbifold decoupling of the partner quarks can be mimicked with the ordinary 
Higgs mechanism in field theory. We use the two-component notation in what follows.
Suppose that at $M_U$ the gauge group becomes ${\rm
SU}(3+N^\prime) \times {\rm SU}(N^\prime)$ (besides ${\rm SU}(2)_L \times
{\rm U}(1)$), where $Q_{Li}$ and $Q_{Li}^\prime$, $u_{Ri}$ and
$u_{Ri}^{\prime}$, $d_{Ri}$ and $d_{Ri}^{\prime}$ are
embedded into fundamental representations of SU$(3+N^\prime)$,
$\Psi_{Qi}$, $\Psi_{{u}i}$, $\Psi_{{d}i}$, respectively,  with
$i$ the generation index. We also introduce anti-fundamental
representations of SU($N^\prime$), $\bar{Q}^{\prime}_{Li}$,
$\bar{u}^{\prime}_{Ri}$, and $\bar{d}^{\prime}_{Ri}$, and a Higgs
field, $\Delta$ which transforms as anti-fundamental and fundamental
representations under SU$(3+N^\prime)$ and SU($N^\prime$), respectively. 
Then, these fields have the following Yukawa terms:\footnote{Note that we have omitted
couplings of the barred fields with the Standard Model Higgs because these couplings 
are absent in the five-dimensional orbifold model.}
\begin{equation}
 {\cal L}_{\rm Yukawa} = \kappa_{Qij} \left(\bar{Q}^{\prime}_{Li} \right)_a 
\Delta^a_{~\alpha} \left(\Psi_{Qj}\right)^\alpha 
+\kappa_{{u}ij} \left(\bar{u}^{\prime}_{Ri} \right)_a 
\Delta^a_{~\alpha} \left(\Psi_{{u}j}\right)^\alpha 
+\kappa_{{d}ij} \left(\bar{d}^{\prime}_{Ri} \right)_a 
\Delta^a_{~\alpha} \left(\Psi_{{d}j}\right)^\alpha 
+{\rm h.c.}~,
\label{eq:decouplingLag}
\end{equation}
where $\alpha = 1, \dots, (3+N^\prime)$ and $a = 1,\dots, N^\prime$. 
We first note that via field redefinitions of $\bar{Q}^{\prime}_{Li}$, $\bar{u}^{\prime}_{Ri}$,
$\bar{d}^{\prime}_{Ri}$, and $\Delta$, it is only possible to make ${\rm arg}({\rm det} \kappa_Q)$, 
${\rm arg}({\rm det} \kappa_u)$, and ${\rm arg}({\rm det} \kappa_d)$ be zero, while the theta 
angle of SU($N^\prime$) is in general nonzero. A zero SU($N^\prime$) theta angle requires 
further UV assumptions (that mimic the CP invariance of the IR brane).

Next, working in this basis, we assume that the Higgs field, $\Delta$ develops the following VEV:
\begin{equation}
\langle \Delta^a_{~\alpha} \rangle = V_\Delta
\begin{pmatrix}
 0&0 &0&1 &0&\ldots &0 \\
 \vdots &\vdots &\vdots & 0& 1 & &\vdots \\
 \vdots & \vdots & \vdots & \vdots & &\ddots &0 \\
0&0&0&0&\ldots &0 &1
\end{pmatrix}
~,
\end{equation}
where $V_\Delta$ can always be taken to be real by using an
SU($3+N^\prime$) gauge transformation. In the  dual CFT picture, this VEV 
corresponds to a condensate of ``techniquarks'' and since the strong ``technicolor'' 
dynamics preserves CP no new phases are introduced. 
This VEV breaks the gauge group into ${\rm SU}(3) \times {\rm SU}(N^\prime)$. 
The upper three components of $\Psi_{{Q,{u},{d}}}$, $Q_L$, $u_R$, and $d_R$, do not
obtain a mass from the VEV, while the lower $N^\prime$ components,
$Q_L^\prime$, $u_R^{\prime }$, $d_R^{\prime }$, form vector-like mass
terms together with $\bar{Q}^{\prime }_{L}$, $\bar{u}^{\prime }_R$, and
$\bar{d}^{\prime }_R$, respectively. Since ${\rm arg}({\rm det}
\kappa_{Q,{u},{d}} ) = 0$, these mass terms do not
contribute to the physical theta term. As a result, we can decouple the
SU$(3+N^\prime)$ partner fields of quarks without spoiling the relation
$\bar{\theta} = \bar{\theta}^\prime$.

\section{Dual interpretation}
\label{sec:dualpicture}

\renewcommand{\theequation}{B.\arabic{equation}}
\setcounter{equation}{0}

The PQ mechanism in four dimensions can also be understood in terms of
the non-dynamical Chern--Simons three-form in QCD and the screening of
the corresponding background ``electric'' field. In this section, we
reinterpret our model setup based on this dual description. However it is
instructive to first consider a simpler two-dimensional model which has
one U(1) gauge field. After that, it will become clear how ${\rm
U}(1)_{\rm PQ}$ is broken, and the axion gets a mass, in the presence of
two gauge fields. The generalization to the four-dimensional dual theory
will then become apparent. 

The standard Schwinger model \cite{Schwinger:1962tp} in two dimensions plus the axion, $a$ has the Lagrangian
\beq
{\cal L} = - \frac{1}{4e^2} F_{\mu\nu}F^{\mu\nu} + \frac{f^2}{2} \left(\partial_\mu a\right)\left(\partial^\mu a\right)
+\frac{1}{2\pi} a\, \varepsilon^{\mu\nu} F_{\mu\nu}\,,
\eeq
where the $\theta$ term has been absorbed in the axion field and $e$ is the U(1) coupling. A crucial point is that the gauge field has no physical propagating degree of freedom in two dimensions, and therefore there is only an instantaneous Coulomb interaction. The only physical degree of freedom is that described by $a$, which is massless at the Lagrangian level (due to the U(1)$_{\rm PQ}$ shift symmetry), but it obtains a mass quantum-mechanically. Simultaneously the Coulomb long-range potential (which grows linearly
at large distances in two dimensions) gets screened. 

First, note that one can always choose the gauge $A_1 \equiv 0$, and then the only remaining component of the 
gauge field is $A_0$, which enters in the Lagrangian without a time derivative,
\beq
{\cal L}_{A_1=0}  = \frac{1}{2e^2} \left(\partial_1 A_0\right)^2  + \frac{f^2}{2} \left(\partial_\mu a\right)\left(\partial^\mu a\right)
- \frac{1}{\pi} a\, \left(\partial_1 A_0\right)\,.
\label{three}
\eeq
In this case, one can immediately eliminate $A_0$ through the classical equation of motion:
\begin{align}
 \label{A0eqn}
A_0
&=
 \frac{e^2}{\pi}\, \partial_1^{-1} a\,,\\[2mm]
 {\cal L}_{A_1=0} 
  &=
  \frac{f^2}{2} \left(\partial_\mu a\right)\left(\partial^\mu a\right) - \frac{e^2}{2\pi^2}a^2\,.
  \label{four}
\end{align}
Hence, the axion mass becomes
\beq
m_a= \frac{e}{\pi\,f}\,.
\eeq
The constraint (\ref{A0eqn}) can also be written as
\beq
\frac 12 \left(\frac 1e\, \partial_1 A_0 -\frac e\pi a
\right)^2\equiv 0\,.
\label{six}
\eeq
Note that $A_0$ is an auxiliary field and does not represent any physical degree of freedom in (\ref{three}), nor does it becomes a degree of freedom after elimination, as in (\ref{A0eqn}).

Next we consider adding a second gauge field, $B_\mu$. The Lagrangian (\ref{three}) now becomes
\beq
{\cal L}_{A_1=0}  = \frac{1}{2e^2} \left[ \left(\partial_1 A_0\right)^2  + \left(\partial_1 B_0\right)^2
\right] 
+ \frac{f^2}{2} \left(\partial_\mu a\right)\left(\partial^\mu a\right)
- \frac{1}{\pi} a\, \left[ \left(\partial_1 A_0\right)+\left(\partial_1 B_0\right)\right] \,.
\label{seven}
\eeq
The most crucial point is that the couplings of the both gauge fields $A_\mu$ and $B_\mu$ are the same.
This is chosen to mimic the unified origin of the separate U(1) fields. The equations of motion for the auxiliary fields 
are now
\beq
A_0
= \frac{e^2}{\pi}\, \partial_1^{-1} a\,,\qquad B_0
= \frac{e^2}{\pi}\, \partial_1^{-1} a\,.
\label{eight}
\eeq
In fact, Eq. (\ref{eight}) has an ambiguity which is sometimes formulated in terms of a
 constant electric field background in the vacuum. Such fields would require electric charges
at the spatial boundary. If one has two distinct U(1) theories 
and assumes two distinct electric charges at the spatial infinities 
for two U(1)'s then, effectively, this would correspond to different ``primordial'' $\theta$'s 
in two U(1)'s. Then, of course, our axion will not be able to ``screen'' both. 
An analogous situation in four dimensions will be to have different $\theta$'s
in SU(3) and SU($N'$) if we ignore their unification. 
We cannot model a unifying non-Abelian group  
in the Schwinger two-dimensional model because, for non-Abelian groups, there is no $\theta$ in two dimensions.
In this case, to model unification we can impose a $Z_2$ symmetry
in the original Lagrangian. Then the boundary conditions at infinity should be $Z_2$ symmetric as well,
implying that the electric background field in the bulk is one and the same for both U(1)'s.

Both auxiliary fields in Eq. (\ref{eight}) are expressed in terms of one and the same physical field $a$, but there is 
no problem with this since $A_\mu$ and $B_\mu$  are auxiliary to begin with. Note that this is not the Higgs mechanism 
in which, if $A_0$ eats up $a$ there is nothing left for $B_0$ to eat up.

Substituting Eq.~(\ref{eight}) in Eq.~(\ref{seven}), the axion mass-squared $m_a^2$ becomes twice as large and Eq.~(\ref{six}) is replaced by
\beq
\frac 12 \left(\frac 1e\, \partial_1 A_0 -\frac e\pi a
\right)^2\equiv 0\,,\qquad \frac 12 \left(\frac 1e\, \partial_1 B_0 -\frac e\pi a
\right)^2\equiv 0.
\label{nine}
\eeq
If we introduce probe electric charges, $Q$  it is not difficult to see that both are screened at distances larger than $1/m_a$.

Finally, it is instructive to comment on the four-dimensional Yang--Mills theory and interpret the axion mechanism with an enlarged color group in the dual formulation introduced in Ref.~\cite{Dvali:2005an}. We will focus on one aspect, namely, the integration constant ambiguities~\cite{gia_argument}. The essence of the effective low-energy dual formulation of Refs.~\cite{Dvali:2005an, gia_argument} is as follows. One introduces a three-form gauge field
\begin{equation}
C_{\alpha\beta\gamma} \propto \varepsilon_{\alpha\beta\gamma\mu}K^\mu~,
\label{D1}
\end{equation}
where $K^\mu$ is the conventional Chern--Simons current. Unlike the Schwinger model, the field $C_{\alpha\beta\gamma}$
is composite. However, in the effective low-energy description one can build the corresponding fully antisymmetric field tensor analogous to $F_{\mu\nu}$ in the Schwinger model, and, add its kinetic term. An analog of Eq. (B.7)
will take the form (symbolically)
\begin{equation}
\partial_{[\mu} C_{\alpha\beta\gamma]} \propto \varepsilon_{\alpha\beta\gamma\mu} a\,.
\label{D2}
\end{equation}
Using the gauge in which $C_{\alpha\beta\gamma}$ with the zero value of one of the subscripts vanishes,
we obviously conclude that $C_{\alpha\beta\gamma}$ is nondynamical (much in  the same way as $A_0$ in (B.7)),
and the solution of Eq. (\ref{D2}) contains an integration constant. 
Note that nondynamical  three-form $C$ fields are sourced by domain walls~\footnote{Strictly speaking, in pure Yang--Mills there are no static domain walls since the vacuum is unique. However, if $N$ is large, there are of order $N$ quasivacua \cite{Witten:1998uka}, which are split from the unique genuine vacuum by a small amount proportional to $N^0$, while the vacuum energy density per se is proportional to $N^2$ (see Refs.~\cite{Shifman:1998if, Gabadadze:2002ff} and references therein). The decay rate of the false vacua is exponentially suppressed.}.
In Refs.~\cite{Dvali:2005an, gia_argument}, it is argued that,
since at low energies we deal with two gauge groups, SU(3) and SU($N^\prime$), there are two independent integration constants. 
This is equivalent to having two distinct $\theta$ angles which would imply, in turn, that a single axion under consideration is 
unable to solve the CP problem.

To our mind the above argument does not take into account that both
low-energy gauge groups are unified at high energies into an SU$(3+N')$ gauge group. This provides us with a unified initial condition for the
$\theta$ angle evolution. In the effective low-energy language of three-form fields this would amount to equality of two integration constants. We do not know at the moment whether this equality is derivable in the effective description \cite{Dvali:2005an, gia_argument} {\em  per se}.

The fact that the overall structure of the $\theta$ parameters (and the associated physical $\theta$ periodicity,
related to the vacuum structure) depends on the topology in the space of fields at all energy-momentum scales, including arbitrarily high, was emphasized in \cite{Shifman:1979if,Dimopoulos:1979qi}. 
In \cite{Dimopoulos:1979qi} it was explicitly noted that in the case of two group factors $G_1$ and $G_2$ (in our model, SU(3) and SU($N^\prime$)) obtained from a unifying group $G$, \textit{i.e.}, $G_1\times G_2\subset G$ at a high scale, the number of independent $\theta$ angles is one rather than two because the  $G_1$ and $G_2$  instantons can be deformed into one another by passing through configurations of arbitrarily large but finite action.  

A very pedagogical example suggested in Ref.~\cite{Dimopoulos:1979qi} is
as follows. Consider the quantum-mechanical problem of a single particle on a circle $S_1$ assuming that the motion on the circle is free. The boundary conditions on the wavefunctions need not be periodic. They can be periodic up to a Bloch phase, provided that one and the same phase enters in the boundary conditions for all wavefunctions. This gives rise to the $\theta$ parameter.

Now, consider instead a particle on a sphere $S_2$ in a potential (defined on $S_2$) such that it has a deep and steep minimum along the sphere's equator. The depth of the trough can be arbitrarily large (but finite) 
so that one might naively say that the low-energy motion of the particle is equivalent to that on $S_1$.

However, this would be the wrong answer, since no matter how high the barrier, the topology of the
configurational space changes, and the Bloch boundary condition is impossible. Tails of the wavefunctions of the 
system ``feel'' that there is a continuous path from an effective $S_1$ to $S_2$.
The $\theta$ angle no longer exists.  Therefore, considering only the low-energy limit tells us nothing about the disappearance of the Bloch boundary condition and the $\theta$ angle.


\newpage
\bibliographystyle{JHEP}
\bibliography{ref}


\end{document}